# Photoluminescence Mapping and Time-Domain Thermo-Photoluminescence for Rapid Imaging and Measurement of Thermal Conductivity of Boron Arsenide

Shuai Yue[1,2,+], Geethal Amila Gamage[3,+], Mohammadjavad Mohebinia[4], David Mayerich[2], Vishal Talari[5], Yu Deng[5], Fei Tian[3], Shen-Yu Dai[6,2], Haoran Sun[3], Viktor G. Hadjiev[7], Wei Zhang[8], Guoying Feng[6], Jonathan Hu[8], Dong Liu[5], Zhiming Wang[1,*], Zhifeng Ren[3, *] and Jiming Bao[2,4,*]

[1]Institute of Fundamental and Frontier Sciences, University of Electronic Science and Technology of China, Chengdu, Sichuan 610054, China
[2]Department of Electronic and Computer Engineering, University of Houston, Houston, Texas 77204, USA
[3]Department of Physics and Texas Center for Superconductivity, University of Houston, Houston, Texas 77204, USA
[4]Materials Science & Engineering, University of Houston, Houston, Texas 77204, USA
[5]Department of Mechanical Engineering, University of Houston, Houston, Texas 77204, USA
[6]College of Electronics & Information Engineering, Sichuan University, Chengdu, Sichuan 610064, China
[7]Texas Center for Superconductivity, University of Houston, Houston, Texas 77204, USA
[8]Department of Electrical and Computer Engineering, Baylor University, Waco, TX 76798, USA

[+]These authors contributed equally to this work.
[*]To whom correspondence should be addressed: zhmwang@uestc.edu.cn, zren@uh.edu, jbao@uh.edu

# Abstract

Cubic boron arsenide (BAs) is attracting greater attention due to the recent experimental demonstration of ultrahigh thermal conductivity $\kappa$ above 1000 W/m·K. However, its bandgap has not been settled and a simple yet effective method to probe its crystal quality is missing. Furthermore, traditional $\kappa$ measurement methods are destructive and time consuming, thus they cannot meet the urgent demand for fast screening of high $\kappa$ materials. After we experimentally established 1.82 eV as the indirect bandgap of BAs and observed room-temperature band-edge photoluminescence, we developed two new optical techniques that can provide rapid and non-destructive characterization of $\kappa$ with little sample preparation: photoluminescence mapping (PL-mapping) and time-domain thermo-photoluminescence (TDTP). PL-mapping provides nearly real-time image of crystal quality and $\kappa$ over mm-sized crystal surfaces; while TDTP allows us to pick up any spot on the sample surface and measure its $\kappa$ using nanosecond laser pulses. These new techniques reveal that the apparent single crystals are not only non-uniform in $\kappa$, but also are made of domains of very distinct $\kappa$. Because PL-mapping and TDTP are based on the band-edge PL and its dependence on temperature, they can be applied to other semiconductors, thus paving the way for rapid identification and development of high-$\kappa$ semiconducting materials.

Cubic boron arsenide (BAs) is one of the most difficult crystals to synthesize due to the inertness of boron (B), high toxicity of arsenic (As), very high melting point of B (~2076 °C), very low sublimation point of As (~615 °C), and the decomposition of BAs into $B_{12}As_2$ at about 920 °C[1, 2, 3]. The interest in BAs increased suddenly in 2013 when first-principles calculations predicted BAs as a highly thermal conductive ($\kappa$) material with a value comparable to that of diamond and much higher than the most commonly used heat-sink materials like copper and silicon carbide[4, 5]. Subsequent intense research led to successful synthesis and verification of the predicted high-$\kappa$ material in 2018 by three groups[6, 7, 8], which opens up new opportunities for more basic research as well as potential applications as a thermal sink material compatible with silicon[9]. However, despite BAs being first studied in the late 50s of the last century[1], its actual bandgap value has not been well settled. The bandgap from latest first principles DFT calculations by several independent groups begins to merge, but still falls in a wide range from 1.7 – 2.1 eV[10, 11, 12, 13]. The bandgap from earlier calculations was either too high[14] or too low[15]. Experimentally, earlier studies put the bandgap down to 1.4-1.5 eV[3, 16, 17]; latest low temperature photoluminescence (PL) combined with hybrid functional calculations brought the bandgap up to 1.78 eV at 0 K[11]. More experiments are needed to confirm the bandgap and a direct experimental measurement of BAs bandgap is still missing.

In addition to the uncertainty in its basic bandgap, current $\kappa$ measurement methods are difficult to meet the demand for rapid screening of $\kappa$ for development of BAs for future device applications. X-ray diffraction (XRD) and transmission electron microscopy (TEM) were used to prove the high crystal quality of BAs in recent demonstrations[6, 7, 8], but it is well known that these techniques are not very sensitive to impurities, point defects, and doping levels, which are important for phonon dynamics and transport[4, 5]. Time-domain thermoreflectance (TDTR) was used by all three groups to determine $\kappa$ of BAs[6, 7, 8], but it is a very specialized technique that requires expensive lasers, sophisticated data collection, analysis and interpretation[18, 19, 20]. TDTR also requires a complicated sample preparation step: the surface of the sample must be polished to be flat enough and is then coated with a metal film that serves as both a laser absorber and a thermal transducer[18, 19, 20]. Other methods, like one-dimensional heat transport method requires even more careful sample preparation[6], and Raman method suffer similar issues as TDTR, so they are not widely employed[6].

In this work, we first experimentally determine the BAs bandgap using optical absorption spectroscopy and then report the observation of room-temperature band edge PL. Based on the relationship of PL with crystal quality and temperature, we develop and demonstrate two new optical techniques that can produce rapid non-intrusive imaging of $\kappa$ over a large region with micrometer spatial resolution with little sample preparation: photoluminescence mapping (PL-mapping) and time-domain thermo-photoluminescence (TDTP). PL-mapping can generate 2D images of $\kappa$ within seconds and even in real-time. TDTP employs two pump-pulse nanosecond lasers to determine $\kappa$ from an arbitrary spot of a sample surface. Since PL-mapping and TDTP use simple lasers, they are accessible by a wide population of researchers, thus will greatly accelerate the pace of discovery, development, and applications of semiconductor-based high $\kappa$ materials.

Bandgap is the most important property of a semiconductor, it is also an essential parameter for theoretical calculation, and forms the basis for optical, electronic as well as thermal properties[21]. To experimentally obtain BAs bandgap, we grew thin single crystals using a recently reported method and chose optical absorption spectroscopy to directly measure its bandgap[22]. Fig. 1a shows UV-Visible absorption spectra of three representative samples at room temperature in Tauc plot as an indirect semiconductor. The nature of indirect bandgap semiconductor is quickly confirmed because despite their differences in absorption at lower energy tails, all of them exhibit the same and well defined absorption edge. The intersections give us a band gap of 1.82 eV, which falls in the range of the latest DFT calculations[10, 11, 12, 13].

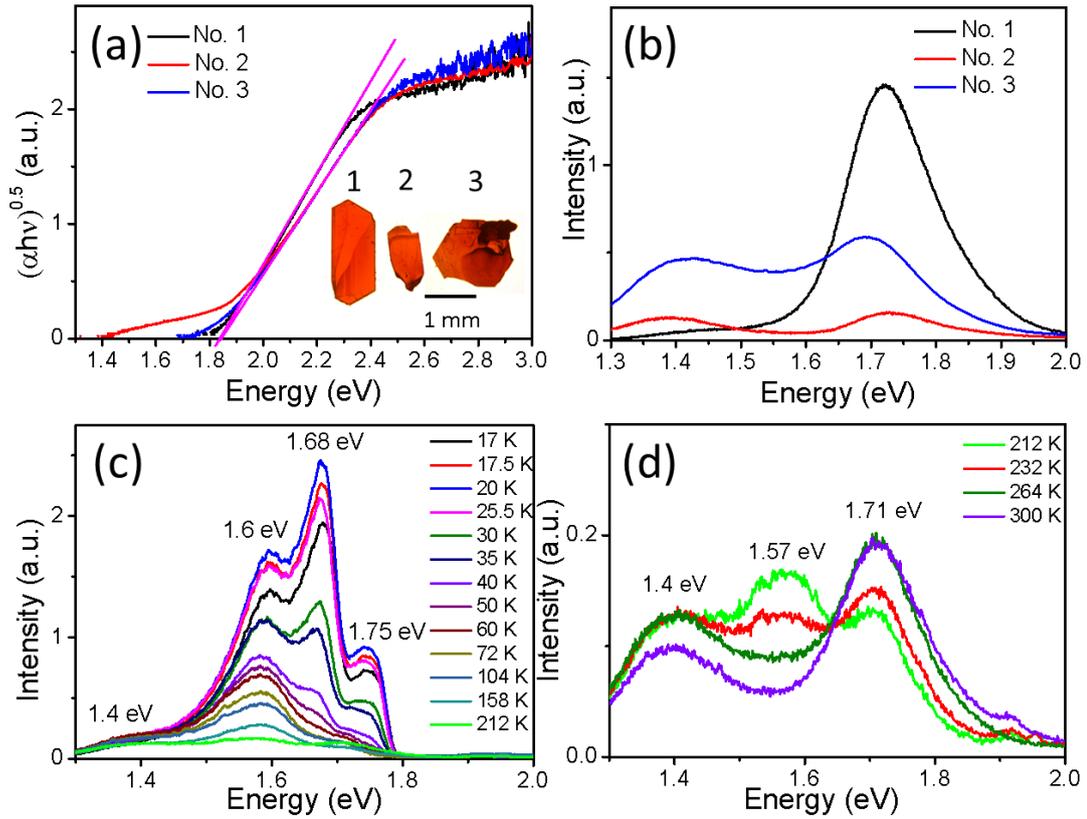

Figure 1. Optical absorption and photoluminescence spectra of BAs crystals. (a) Indirect bandgap Tauc plots of UV-Visible absorption spectra of three BAs crystals. Inset: optical transmission images. (b) Room temperature PL spectra of the samples. (c-d) Evolution of PL of a BAs as a function of temperature (c) from 17 to 212 K and (d) from 212 to 300 K.

Compared to UV-Vis, PL is more sensitive to the crystal quality, defects and doping levels[21]. Since the recent high $\kappa$ demonstrations are actually successful stories of material synthesis of high quality BAs crystals, we performed PL spectroscopy at room temperature although earlier BAs samples exhibited no PL at all[23]. Despite being an indirect semiconductor, the samples reported here show very strong PL as displayed in Figure 1b, indicating a much better crystal quality. The spectra feature a high energy peak at 1.72 eV and a relatively weak peak at 1.4 eV. A close comparison between the PL and absorption spectra reveals a close relationship: Sample #1 with a weak absorption below the absorption edge exhibits a dominant PL peak at 1.72 eV, while Samples #2 and #3 with stronger absorption at the absorption tails also exhibit a strong PL at the same energy. We conclude that the 1.72 eV peak is the band edge associated photoluminescence. The 1.40 eV peak comes from sub-bandgap defect states, which is believed to be responsible for the observed bandgap values in earlier reports[3, 16, 17].

The room temperature PL in Fig. 1b has less spectral features than that of recently observed PL at 10 K which was claimed to be dominated by defect-bound excitons, donor-acceptor transitions, and free exciton transition observed only under a strong excitation[11]. To complete the PL spectra comparison, we varied the temperature and monitored the evolution of PL. As can be seen from Fig. 1c-d, the PL at 17 K resembled the reported features in Ref. [11] very well and confirms the peak at 1.72 eV as due to band edge free exciton transitions. As temperature increases, the intensity of all three major peaks decreases and begins to merge. At ~160 K, the bound exciton peak is merged into the free exciton peak. At ~260 K, the donor-acceptor transition peak disappears. The disappearance of these peaks at higher temperatures is due to thermal excitation of loosely bound electrons and holes. In contrast, the intensity of the defect state at 1.40 eV changes very little, proving that it is from a different origin.

Based on well-established relationship between PL and crystal quality and the relationship between crystal quality and thermal conductivity $\kappa$ proved by recent demonstrations and theories[4, 5, 6, 7, 8, 21], we speculate that a strong PL must lead to a high $\kappa$. To make use of these relationships for thermal conductivity measurement, we take one step further and utilize confocal PL imaging to evaluate $\kappa$ of large BAs crystals. Figure 2 shows the optical pictures and PL-mappings of three BAs crystals at room temperature. The PL-mapping was performed with a custom confocal laser-scanning fluorescent microscope using a 532-nm 40-mW pulsed diode laser to raster scan the sample surface and a GaAsP photo multiplier tube (PMT) with sensitivity in the range of 300 – 720 nm. A 635-nm long-pass filter was used to block PL excitation. Depending on the mapping size and resolution, PL-mappings of millimeter sized samples takes seconds or less to complete, and a real time PL-mapping is also easy to achieve. Despite apparent single crystal sample, PL-mappings reveal a non-uniform intensity (Fig. 2b and d) across the smooth surfaces (Fig. 2a and c), which indicates a non-uniformity of crystal quality in the same crystalline surface. PL-mapping further reveals very different domain patterns (Fig. 2f and h) on the opposite surfaces of the same BAs crystal (Fig. 2e and g). The measured $\kappa$ values shown in Fig. 2d will be explained in details in the next paragraph.

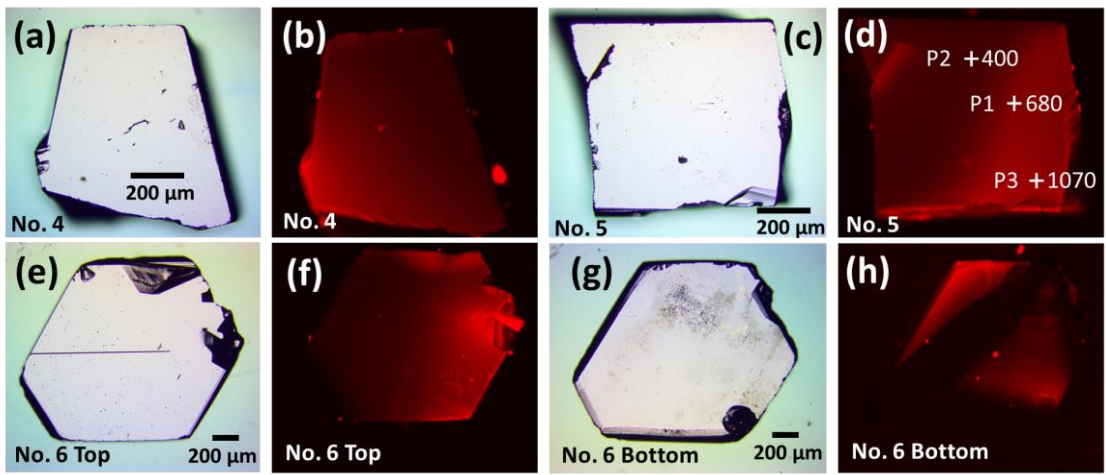

Figure 2. PL-mapping of three BAs samples. Optical images (a, c) and PL-mappings (b, d) of samples No. 4 and No. 5. The selected spots P1-3 are marked by "+" along with their $\kappa$ measured with TDTP. Optical images (e, g) and PL-mappings (f, h) of opposite surfaces of sample No. 6.

To obtain a quantitative $\kappa$ and experimentally verify the relationship between $\kappa$ and PL, we further developed an optical technique that is convenient and compatible with PL and requires no extra sample preparation. Figure 3a shows the schematic of pump-probe time-domain thermo-photoluminescence (TDTP). In parallel to TDTR[18, 19, 20], TDTP uses a pump pulse (527 nm, 140 ns) to generate localized heat in a sample, and a weak time-delayed probe pulse (527 nm, 100 ns) to monitor the temperature and heat transfer dynamics. $\kappa$ is obtained by fitting the measured temperature data to the known analytical/numerical solutions. However, there are very distinct differences between TDTP and TDTR. Instead of ultrafast femtosecond or picosecond lasers used in TDTR, TDTP uses nanosecond lasers which are much cheaper and accessible. The local heat in the material is directly generated by the material's optical absorption of pump pulse, so there is no metal film coated on the surface as in TDTR. The temperature of the sample is obtained by its temperature-dependent PL. To eliminate the PL excited by the pump, the probe pulses are modulated and the PL is detected at the probe modulation frequency.

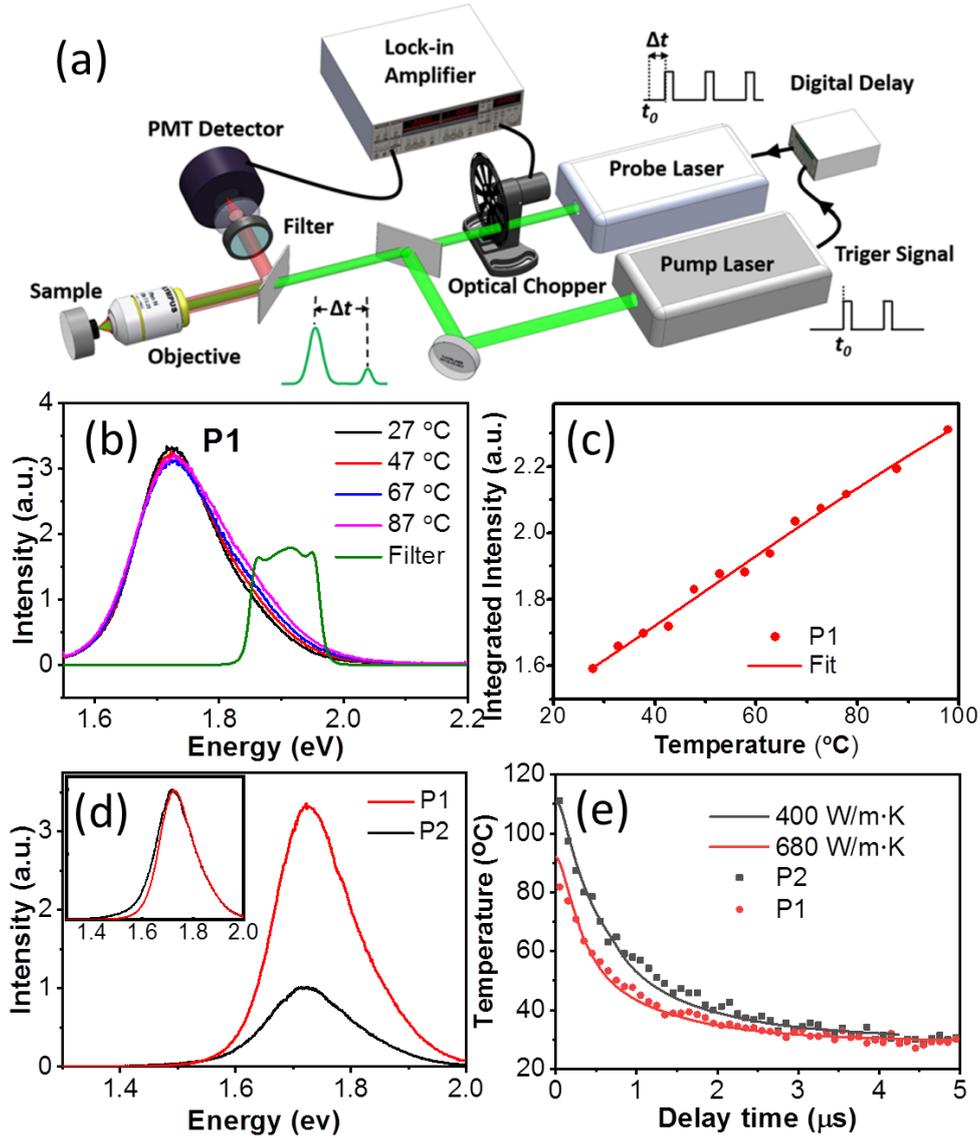

Figure 3. Time-domain thermo-photoluminescence (TDTP). (a) Schematic of TDTP. (b) Temperature-dependent photoluminescence spectrum of the spot P1 in Fig. 2c-d, and the transmission spectrum of a filter used to select the spectral window for TDTP. (c) PL intensity integrated over the transmission window as a function of temperature. (d) Photoluminescence spectra of spots P1 and P2 as shown in Fig. 2c-d. (e) TDTP traces for the P1 and P2, and simulated curves.

Temperature affects the PL spectrum of a material in many ways. Changes in this property can be used to measure a sample's temperature[23]. Figure 3b shows an example of temperature PL shift. The spot P1 was chosen from sample No. 5 as shown in Fig. 2c-d. The spectrum from spot P1 is shown in Fig. 3b. As the temperature increases, the PL peak intensity remains the same, but the short wavelength spectral region becomes enhanced (Fig. 3b). We believe this is due to increased rate of optical transitions of thermally excited electrons and holes that involve absorption of phonons around the indirect bandgap[21, 24]. To utilize this temperature-dependent PL spectrum as a sensitive temperature marker in TDTP, we use a band-pass optical filter to select a short wavelength band and then use a fast PMT to measure the integrated intensity over that band. The spectrum of that band is shown in Fig. 3b, and Fig. 3c shows an almost linear dependence of the integrated PL intensity as a function of temperature; this same curve will be used to determine the sample temperature.

To make a good comparison, we chose another spot P2 from the PL-mapping in Fig. 2c-d. Figure 3d shows their PL spectra, note that the spot with stronger PL also has a narrower PL line width. Figure 3e shows their TDTP temperature evolutions. It is very clear that the spot P1 has a lower temperature, indicating a higher $\kappa$. The temperature difference between the two spots is the highest right after the excitation of pump pulse, and both the temperatures and their difference drop quickly in a few μs due to fast heat diffusion. Using any commercial heat transport software, we can fit TDTP curves with $\kappa$ as a parameter. The simulation results with SIMULIA ABAQUS or COMSOL shown in Figure 3e give us a room-temperature $\kappa$ of 680 W/m·K for the stronger PL spot P1 and 400 W/m·K for the weaker PL spot P2 (their spot locations are shown in Fig. 2d). The highest $\kappa$ corresponding to the strongest PL spot P3 in Fig. 2d reaches 1070 W/m·K, in agreement with the recent demonstrations by TDTR[6-8].

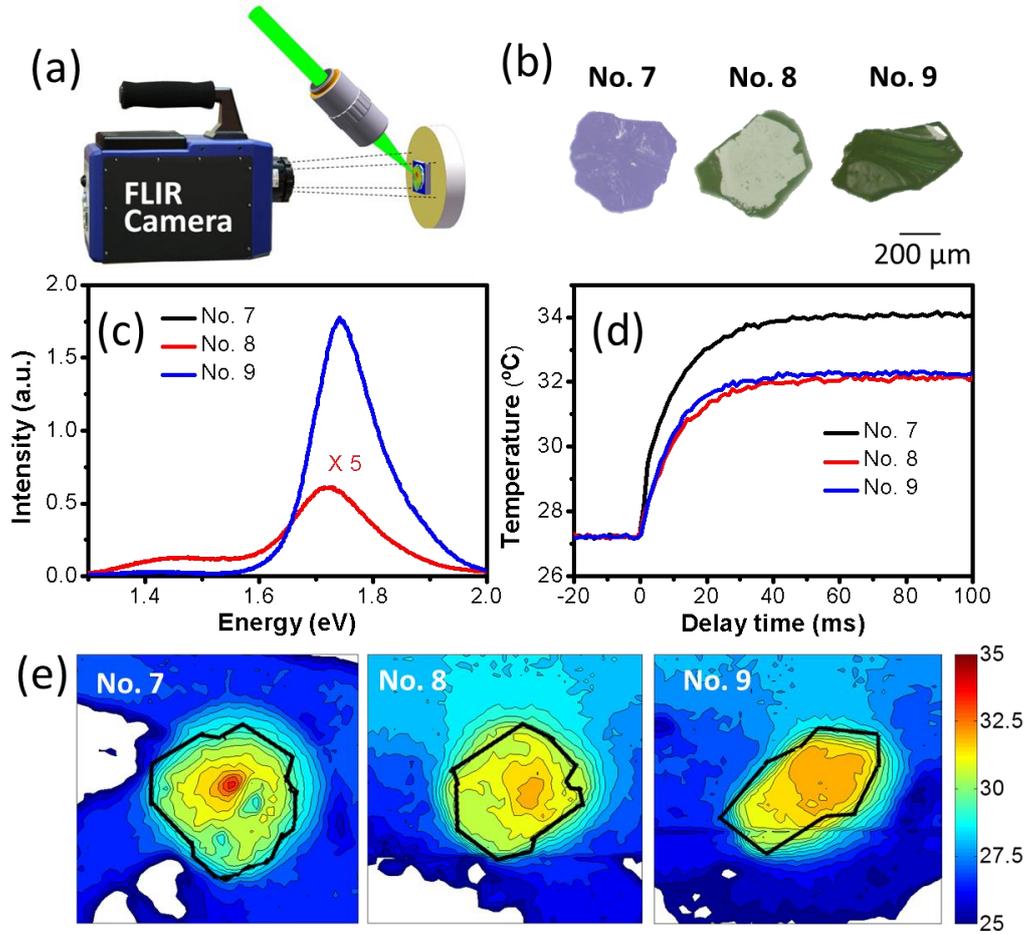

Figure 4. Comparison of $\kappa$ using laser induced heating and infrared thermal imaging. (a) Schematic of experimental setup. (b) Optical images of three samples. (c) Their PL spectra. (d) Temperature evolution of laser spots on the three samples. Laser was turned on at time zero. (e) Contour map of surface temperatures of the three samples at 40 ms. The heating power of a 532-nm continuous wave (CW) laser is 70 mW.

Having shown the power of PL-mapping and TDTP, we note that some region of the sample No. 6 in Fig. 2g-h exhibited weak or no PL even though it was grown by the same recently reported method[22]. In fact, our early BAs samples before this work also exhibited no PL[25]. Since TDTP is not applicable to samples without PL, we developed another technique that requires little sample preparation as PL-mapping and TDTP to quickly prove that samples with no PL have a much lower $\kappa$. Fig. 4a shows the experimental setup. Here a continuous wave (CW) laser is used to induce local heating, and an infrared thermal camera is used to monitor the temperature evolution directly. To have a good comparison, we chose three samples with similar sizes and thickness. Fig. 4b shows their optical images, and their PL spectra are shown in Fig. 3c. Sample No. 7 was from our early batch that does no show PL[25]. The other two were recently grown[22], but sample No. 9 has a much stronger PL than No. 8. Figure 4d shows that, for all three samples, the temperature at the laser focus point increases quickly initially and eventually reaches a plateau. Fig. 4e shows snapshots of surface temperatures for the three samples at 40 ms. Simulation with COMSOL gives us thermal conductivity $\kappa$ of 160 W/m·K for sample No. 7 (Fig. S3), confirming that the sample with no PL has lower $\kappa$ due to poor crystal quality. However, the technique fails to distinguish samples No. 8 and No. 9, both of them reached similar temperatures despite their huge difference in PL intensity. Apparently this failure is due to slow mechanical switching of the CW laser and millisecond slow response of the thermal camera.

The failure of the slow thermal camera for high $\kappa$ materials reminds us of the importance of detection speed; however, it does not mean necessarily that the faster detection speed results in better data. For most heat transport study as demonstrated by TDTP, nanosecond resolution is sufficient and even perfect from the basic physical consideration, because this is the time for photo-excited carriers to fully relax and reach thermal equilibrium with lattice, so that the actual temperature of lattices can be accurately probed by nanosecond PL. This is also true from practical point of view, because for samples with thickness on the order of micrometers, the thermal relaxation time is on the order of nanosecond. Ultrafast lasers are required in TDTR mainly because of ultrafast thermal relaxation of the thin metal film thermal transducer[18, 19, 20]. This thermal relaxation time certainly cannot limit the application of TDTP. For thin films and even atomically thin 2D materials, we can suspend them or place them on a good thermal insulating substrate to eliminate out of plane heat loss. Thus, like PL and PL-mapping, TDTP can handle most samples regardless of their size and thickness.

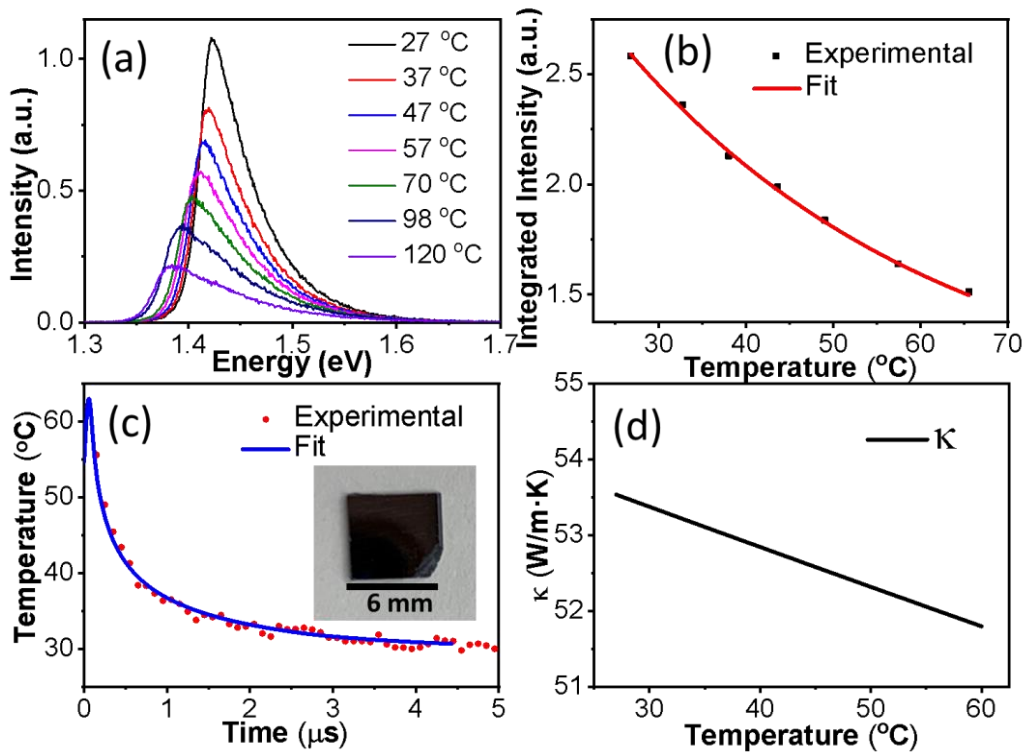

Figure 5. Thermal conductivity $\kappa$ of GaAs measured by TDTP. (a) Temperature-dependent photoluminescence spectrum. (b) Integrated PL intensity as a function of temperature. (c) TDTP trace and simulation. (d) Temperature-dependent $\kappa$ from simulation in (c).

To prove the reliability and accuracy of TDTP, we choose high quality GaAs as a reference to measure its known $\kappa$ using TDTP. Because its bandgap is lower than that of BAs, the same 527-nm laser will also work for GaAs. Figure 4a shows the temperature-dependent PL spectrum of GaAs. At higher temperatures the PL of GaAs is dominated by the direct band-to-band radiative recombination with energy that follows the temperature dependence of the bandgap. In this case, the total integrated PL intensity is an excellent temperature marker. Fig. 4b shows the integrated PL intensity as a function of temperature measured by the same PMT as for Bas. Based on the TDTP data and simulation in Fig. 4c, we obtained a temperature-dependent $\kappa$ of GaAs, as shown in Fig. 4d; it agrees well with published values of GaAs[26], which firmly proves that our TDTP method is reliable for measuring the $\kappa$ of BAs.

Having successfully demonstrated TDTP, we want to stress that TDTP still needs a lot of improvement. Like many other optical techniques such as TDTR, there is a lot of room for improvement and refinement that has been going on for many years. For example, the laser pulse width can be reduced to achieve higher time resolution. The laser pulse shape and power can be adjusted to make heat generation more accurate and controllable. Like PL-mapping, TDTP can be optimized and automated to speed up data acquisition and reduce signal fluctuations. In this demonstration of concept, the experiment was not optimized and each data was obtained point by point manually; as a result, both temperature calibration and TDTP curves are not smooth, and the simulation fits are also not perfect. Based on the quality of BAs TDTP data as well as the comparison of $\kappa$ using GaAs, we believe our measured $\kappa$ of BAs has a 10-15% uncertainty. The essence of combining PL-mapping and TDTP is that it can quickly screen the crystal quality uniformity and find the highest $\kappa$ value so that researchers can further refine the growth method to achieve crystals with uniformly high $\kappa$ for any possible applications.

In summary, we have experimentally determined the bandgap of BAs and observed its band-edge photoluminescence at room temperature as a measure of crystal quality. We subsequently developed and demonstrated PL-mapping and TDTP as quick and reliable screening methods to measure $\kappa$ values of semiconductors using nanosecond lasers with little sample preparation. Our PL-mapping technique revealed non-uniformity of $\kappa$ and different $\kappa$ domains in apparent single crystals BAs, which allowed us to identify region with highest $\kappa$ and guide us to improve the growth techniques. TDTP accurately measures the $\kappa$ of selected spots from the PL-mapping images. As rapid non-contact optical techniques with little sample preparation, PL-mapping and TDTP will greatly accelerate the search for high $\kappa$ semiconductors for device applications.

**Acknowledgements**

Work in Z.F. Ren's lab was supported by the U.S. Office of Naval Research under MURI Grant N00014-16-1-2436. J.M. Bao acknowledges support from Welch Foundation (E-1728) and National Science Foundation (EEC-1530753).